\documentclass[aps,prl,twocolumn,groupedaddress]{revtex4}

\usepackage{graphics}

\begin{document}


\title{Direct observation of charge inversion by multivalent ions as a universal electrostatic phenomenon}


\author{K. Besteman, M. A. G. Zevenbergen, H. A. Heering, and S. G. Lemay}
\affiliation{Kavli Institute of Nanoscience, Delft University of
Technology, 2628 CJ Delft, The Netherlands}


\date{\today}

\begin{abstract}
We have directly observed reversal of the polarity of charged
surfaces in water upon the addition of tri- and quadrivalent ions
using atomic force microscopy. The bulk concentration of
multivalent ions at which charge inversion reversibly occurs
depends only very weakly on the chemical composition, surface
structure, size and lipophilicity of the ions, but is dominated by
their valence. These results support the theoretical proposal that
spatial correlations between ions are the driving mechanism behind
charge inversion.
\end{abstract}


\maketitle

Understanding screening due to mobile ions in liquid is a key
theme of such diverse fields as polymer physics, nanofluidics,
colloid science and molecular biophysics. Screening by multivalent
ions in particular results in several counter-intuitive phenomena,
for example attraction between like-charged macromolecules such as
DNA \cite{bloomfield} and actin filaments \cite{wongactin}.
Similarly, the electrophoretic mobility of charged colloids
reverses sign upon introducing a sufficient concentration of
multivalent ions in solution \cite{james,molinaelec}, a phenomenon
known as charge inversion.

The conventional paradigm for describing screening in liquid
divides the screening ions into two components: (1) the so-called
Stern layer, consisting of ions confined to the surface, and (2) a
diffuse component described by the Poisson-Boltzmann (PB) equation
that decays exponentially with distance far from the charged
object. Charge inversion can be accounted for by introducing a
''chemical'' binding constant that reduces the free energy of
multivalent ions situated in the Stern layer, reflecting an
assumed specific interaction between these ions and the surface
being screened. This chemical binding constant is expected to
depend on properties of the ions such as their size, chemical
composition, surface structure, lipophilicity and valence. While
this approach has been successful in \textit{describing}
experimental data \cite{james,pashleychinv,agashe,vithcuafm}, it
usually provides little insight into the underlying binding
mechanism and lacks significant predictive power.

A universal mechanism for charge inversion based predominantly on
electrostatic interactions has been proposed \cite{reviews}. It
was noted that the predicted chemical potential of the Stern layer
can be significantly lowered if spatial correlations between
discrete ions are accounted for. At room temperature, the loss of
entropy entailed by the formation of a highly correlated ionic
system is substantial. For multivalent counterions and
sufficiently high surface charge densities, however, this is more
than compensated by the corresponding gain in electrostatic
energy, leading to charge inversion \cite{shklpre}. To date, these
theories have remained untested by experiments.

Here we present direct measurements of charge inversion and its
dependence on the properties of the screening ions. Using an
atomic force microscope (AFM), we measured the force between two
oppositely charged surfaces. This approach circumvents the main
limitations of previous measurements, namely, reliance on
modelling of hydrodynamic effects \cite{james,molinaelec} and the
need to disentangle phenomena at two similarly-charged surfaces
\cite{pashleychinv,vithcuafm}. We observe that in the presence of
a sufficiently high concentration of tri- and quadrivalent ions,
the force reversibly changes sign. The bulk concentration at which
charge inversion occurs, the so-called charge-inversion
concentration $c_0$, depends almost exclusively on the valence of
the ions, consistent with the universal predictions of the
ion-correlation theories.

\begin{figure}
\includegraphics{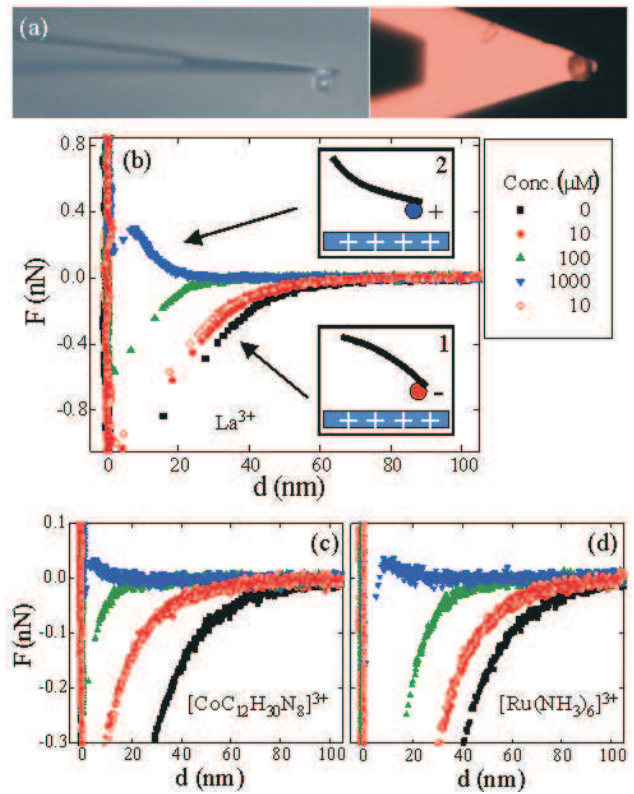}
\caption{\label{Figure1} (color) (a) Optical microscope images of
the side (left) and top (right) of a cantilever with a silica
sphere. Force versus separation measurements in different
concentrations of (b) LaCl$_3$, (c) CoC$_{12}$H$_{30}$N$_8$Cl$_3$
(c) and (d) Ru(NH$_3$)$_6$Cl$_3$. Insets illustrate schematically
the attractive (1) and repulsive (2) forces between the silica
bead and the amine-terminated surface. The legend applies to all
three graphs.}
\end{figure}

Positively charged amine-terminated surfaces were prepared under
argon atmosphere by immersing silicon wafers with 200-500 nm
thermally-grown oxide in a 0.1\% solution of
1-trichlorosilyl-11-cyanoundecane (Gelest) in toluene for 30
minutes, then in a 20 \% solution of Red Al (Sigma-Aldrich) in
toluene for 5 hours. Negatively charged surfaces were prepared by
gluing 10~$\mu$m diameter silica spheres (G. Kisker Gbr) with
epoxy resin to AFM cantilevers (ThermoMicroscope Microlevers)
using a method similar to that of Ducker et al \cite{duckerbead},
as illustrated in Fig.~\ref{Figure1}(a).

Using a Digital Instrument NanoScope IV AFM, force spectroscopy
measurements were performed yielding the force $F$ on the silica
bead versus the bead-surface separation $d$ \cite{duckerbead}. The
spring constant of the cantilevers was 0.03 N/m, as given by the
manufacturer. Care was taken to minimize the scattering of light
from the surface so as to eliminate interference effects.

At separations $d$ greater than the Debye length $\lambda$ of the
solution, the force decays exponentially with $d$:
\begin{equation}
\label{eq:f} F = F_0 \exp (-d/\lambda), \qquad d > \lambda.
\end{equation}
The parameter $F_0$ is proportional to the so-called renormalized
surface charge densities of both the silica bead and the
amine-terminated surface, $\sigma_b^\star$ and $\sigma_s^\star$
respectively. The values of $\sigma_{b,s}^\star$ are related by
the PB equation to the net surface charge densities $\sigma_{b}$
and $\sigma_{s}$ (including both the bare surface charge and the
charge in the Stern layer). At low net surface charge densities
$|\sigma_{b,s}| < \sigma_{\mathrm{max}}$, the renormalized charge
densities are simply equal to the net charge densities:
$\sigma_{b,s}^\star = \sigma_{b,s}$. Here $\sigma_{\mathrm{max}} =
4~kT\epsilon/e\lambda$, where $k$ is the Boltzmann constant, $T$
is the temperature, $\epsilon$ is the dielectric constant of water
and $-e$ is the electron charge. At higher net charge densities,
$\sigma_{b,s}^\star$ saturates at $\sigma_{\mathrm{max}}$. Because
we use oppositely charged surfaces and $Z$:1 electrolytes, where
$Z$ is the valence of the multivalent ions, only one of the
surfaces is affected by the introduction of multivalent ions. The
other surface thus plays the role of a constant probe. Near charge
inversion, $F_0$ is thus proportional to the net surface charge
density of the surface being screened by multivalent ions,
$\sigma_{b}$ or $\sigma_{s}$, and the sign of the force yields
unambiguously the polarity of this net surface charge.

For $d\lesssim \lambda$, the PB equation predicts a more
complicated form than Eq.~(\ref{eq:f}). In addition, van der Waals
forces, regulation of the surface charge and depletion forces can
become important. We therefore concentrate our analysis on the
regime $d
> \lambda$.

Three positive trivalent ions, Lanthanum La$^{3+}$, ruthenium(III)
hexammine [Ru(NH$_3$)$_6$]$^{3+}$ and cobalt(III) sepulchrate
[CoC$_{12}$H$_{30}$N$_8$]$^{3+}$ were investigated. La$^{3+}$ is
an elemental metal ion with a first hydration shell consisting of
8--9 water molecules (radius $r$ of the complex 398~pm
\cite{note1, shannon, marcus, burgess}). [Ru(NH$_3$)$_6$]$^{3+}$
contains a Ru(III) core with six NH$_3$ groups around it
($r=364$~pm \cite{note1,shannon,marcus}).
[CoC$_{12}$H$_{30}$N$_8$]$^{3+}$ is a caged cobalt complex with
CH$_2$ groups exposed to the water environment ($r=445$~pm
\cite{bacchi}), making it less hydrophillic than the other two.

Figure~\ref{Figure1} shows the measured force-distance relation
$F(d)$ as a function of multivalent ion concentration $c$ for the
multivalent salts LaCl$_3$ (b), CoC$_{12}$H$_{30}$N$_8$Cl$_3$ (c)
and Ru(NH$_3$)$_6$Cl$_3$ (d). A force measurement with only
supporting electrolyte (LaCl$_3$: \cite{suppelec1a},
CoC$_{12}$H$_{30}$N$_8$Cl$_3$ and Ru(NH$_3$)$_6$Cl$_3$:
\cite{suppelec1bc}) was first performed (black squares), showing
an attractive interaction between the surfaces. Solutions with
increasing concentrations of multivalent ions in the monovalent
supporting electrolyte were then pumped through the AFM fluid cell
of 50 $\mu$l volume at a rate 0.15--0.2 ml/min for at least 5
minutes per solution. This allowed the surface to equilibrate with
the electrolyte and insured that $c$ was not influenced when large
numbers of ions screened the surface. Consecutive measurements of
$F(d)$ at multivalent ion concentrations $c$ = 10~$\mu$M,
100~$\mu$M and 1~mM are shown in Fig.~\ref{Figure1}. At the end of
the experiment, the measurement with $c$ = 10~$\mu$M was repeated
(red open circles). The CoC$_{12}$H$_{30}$N$_8$Cl$_3$ and
Ru(NH$_3$)$_6$Cl$_3$ measurements were carried out consecutively
using the same silica bead.

We interpret these observations as follows. The positive
multivalent ions adsorb on the negative silica bead, reducing
$\sigma_b$ and thus the magnitude of the force. Near 1~mM, the
screening charge in the Stern layer overcompensates for the bare
surface charge; $\sigma_b$ becomes positive and the force becomes
repulsive. The last measurement with $c$ = 10~$\mu$M, which shows
a recovery to the force measured at the beginning of the
experiment, indicates that charge inversion reflects reversible
equilibrium between the surface and the bulk electrolyte.

To further compare the charge-inversion concentration of the same
surface with different multivalent ions, the force for $d >
\lambda$ was fitted to Eq.~(\ref{eq:f}). Because it is difficult
to accurately fit the Debye length $\lambda$ when the force is
very small, its value was fitted for the curve with $c=0$ and
corrected using the standard expression for $\lambda$ for the
cases $c>0$.

\begin{figure}
\includegraphics{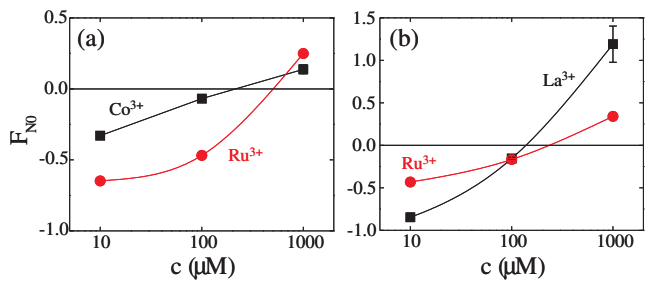}
\caption{\label{Figure2} Normalized force extrapolated to zero
separation obtained from fits to Eq.~(\ref{eq:f}), versus
multivalent ion concentration $c$ for (a)
CoC$_{12}$H$_{30}$N$_8$Cl$_3$ (squares) and Ru(NH$_3$)$_6$Cl$_3$
(circles) and for (b) La$^{3+}$ (squares) and Ru(NH$_3$)$_6$Cl$_3$
(circles). In each case the data were obtained consecutively using
the same silica bead. Lines are guides to the eye.}
\end{figure}

Figure~\ref{Figure2}(a) shows the fitted normalized force
extrapolated to zero separation, $F_{N0}(c) = F_0(c)/F_0(0)$, for
the [CoC$_{12}$H$_{30}$N$_8$]$^{3+}$ and [Ru(NH$_3$)$_6$]$^{3+}$
data of Fig.~\ref{Figure1}(c,d). Similarly, Fig.~\ref{Figure2}(b)
shows $F_{N0}(c)$ for consecutive measurements using the same
silica bead on La$^{3+}$ (data from Fig.~\ref{Figure1}(b)) and
[Ru(NH$_3$)$_6$]$^{3+}$ ($F(d)$ curves not shown). We estimate the
charge-inversion concentration $c_0$ by linearly interpolating
between the data points immediately above and below $F_{N}=0$ on
the lin-log scale. In both sets of measurements, the observed
values of $c_0$ differ by a factor $\sim2$. More generally, we
find that the charge-inversion concentrations of silica for the
three chemically different trivalent ions La$^{3+}$,
[Ru(NH$_3$)$_6$]$^{3+}$ and [CoC$_{12}$H$_{30}$N$_8$]$^{3+}$
differ by at most a factor of 2.1, as summarized in
Table~\ref{Table1}. This is comparable to the variation observed
between measurements for the same ion and pH using different,
nominally identical beads and surfaces. Although the
charge-inversion concentrations of the three positive trivalent
ions are similar, there are differences in the observed $F(d)$
curves. In particular, La$^{3+}$ is less effective in reducing the
absolute force at low concentrations, but the magnitude of the
force for $c \gg c_0$ is largest for La$^{3+}$.

\begin{figure}
\includegraphics{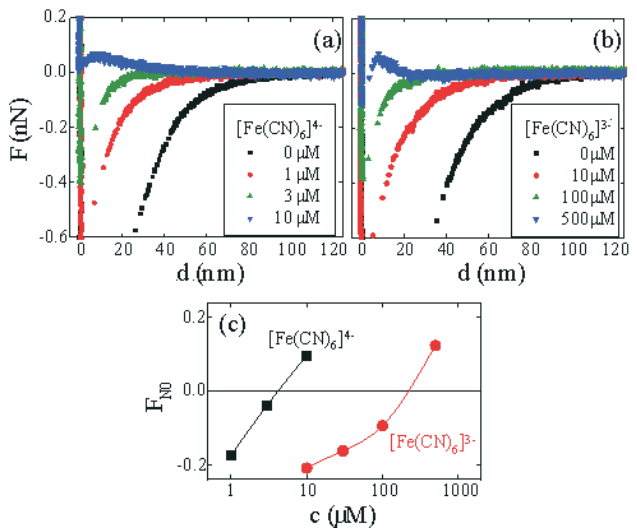}
\caption{\label{Figure3} (color) Force versus separation
measurements in different concentrations of (a) K$_4$Fe(CN)$_6$
and (b) K$_3$Fe(CN)$_6$. (c) Normalized force at zero separation
versus multivalent ion concentration $c$ for K$_4$Fe(CN)$_6$
(squares) and K$_3$Fe(CN)$_6$ (circles). Lines are guides to the
eye.}
\end{figure}

Figure~\ref{Figure3} shows measurements where the same
amine-terminated surface was consecutively charge inverted by a
molecule in two different charge states, iron(II) hexacyanide
[Fe(CN)$_6$]$^{4-}$ ($r=443$~pm) and iron(III) hexacyanide
[Fe(CN)$_6$]$^{3-}$ ($r=437$~pm) \cite{note1,shannon,marcus},
ensuring that essentially the only difference between the two
measurements is the valence of the ions \cite{suppelec3}.
Figure~\ref{Figure3}(c) shows the $F_{N0}$ for both ions as a
function of the concentration. The charge-inversion concentrations
for the two ions differ by a factor $\sim50$.

Measurements using [Fe(CN)$_6$]$^{4-}$ and ruthenium(II)
hexacyanide [Ru(CN)$_6$]$^{4-}$ ($r=456$~pm
\cite{note1,shannon,marcus,note2}), 
two ions with nearly identical chemical groups exposed to solution
and differing only by their core atom, gave nearly identical
$F(d)$ curves at all concentrations.

Two divalent ions, calcium Ca$^{2+}$ and magnesium Mg$^{2+}$
(radii of 388 and 348~pm, respectively \cite{note1,shannon,marcus,
burgess})
did not show charge inversion at a concentration of 1~mM on a
silica bead that showed charge inversion at 1~mM La$^{3+}$. Thus
divalent ions, if they can charge invert a silica bead at all, do
so at higher concentrations than trivalent ions. Concentrations
higher than 1mM were not investigated in this study because the
Debye length then becomes so short that effects such as van der
Waals forces mask the electrostatic interaction between the
surfaces.

Additional experiments were performed with positively charged
surfaces made by chemically modifying a silicon dioxide surface
with 3-aminopropyltriethoxysilane (APTES) and by adsorbing
poly-L-lysine on mica. Key results are summarized in
Table~\ref{Table1}

\begin{table*}
\caption{\label{Table1} Summary of measurements in which the same
surface was charge inverted by two different ions.}
\begin{ruledtabular}
\begin{tabular}{lccccccc}
surface & probe & supp. elect. & ion(1) & ion(2) & $c_0^{(1)}$
($\mu$M) & $c_0^{(2)}$ ($\mu$M) & $c_0^{(high)}/c_0^{(low)}$
\\ \hline
APTES & silica bead & \cite{suppelecmes} &
$\lbrack$Fe(CN)$_6$$\rbrack$$^{4-}$ &
$\lbrack$Fe(CN)$_6$$\rbrack$$^{3-}$ & 13 & 170 & 13 \\
chlorosilane & silica bead & \cite{suppelec3} &
$\lbrack$Fe(CN)$_6$$\rbrack$$^{4-}$ &
$\lbrack$Fe(CN)$_6$$\rbrack$$^{3-}$ & 4 & 200 & 50 \\
APTES & silica bead & \cite{suppelecmes} &
$\lbrack$Ru(CN)$_6$$\rbrack$$^{4-}$ &
$\lbrack$Fe(CN)$_6$$\rbrack$$^{4-}$ & 11 & 13 & 1.2 \\
silica bead & APTES & \cite{suppelecmes} & La$^{3+}$ &
$\lbrack$Ru(NH$_3$)$\rbrack$$^{3+}$ & 560 & 730 & 1.3 \\
silica bead & poly-L-lysine & \cite{suppelec1a} &
$\lbrack$CoC$_{12}$H$_{30}$N$_8$$\rbrack$$^{3+}$ & La$^{3+}$ & 190 & 120 & 1.6 \\
silica bead & poly-L-lysine & \cite{suppelec1a} &
$\lbrack$CoC$_{12}$H$_{30}$N$_8$$\rbrack$$^{3+}$ & La$^{3+}$ & 170 & 180 & 1.1 \\
silica bead & chlorosilane & \cite{suppelec1a} & La$^{3+}$ & $\lbrack$Ru(NH$_3$)$\rbrack$$^{3+}$ & 130 & 210 & 1.6 \\
silica bead & chlorosilane & \cite{suppelec1bc} & $\lbrack$CoC$_{12}$H$_{30}$N$_8$$\rbrack$$^{3+}$ & $\lbrack$Ru(NH$_3$)$\rbrack$$^{3+}$ & 210 & 450 & 2.1 \\
poly-L-lysine & silica bead & \cite{suppelecmes} &
$\lbrack$Ru(CN)$_6$$\rbrack$$^{4-}$ & & 22 \\
\end{tabular}
\end{ruledtabular}
\end{table*}

In terms of a chemical binding description, our measurements
indicate that the binding constants for La$^{3+}$,
[Ru(NH$_3$)$_6$]$^{3+}$ and [CoC$_{12}$H$_{30}$N$_8$]$^{3+}$ on
silica differ by at most a factor $\sim2$, despite the fact that
these ions have significantly different chemical composition,
surface structure, size and lipophilicity. The binding constant
differs at least 10-fold for the same molecule in two different
charge states on amine-terminated surfaces. These observations
strongly suggest that specific chemical interactions are not
responsible for charge inversion in our measurements and that the
mechanism for adsorption is predominantly electrostatic.

We compare our results with ion-correlation theories using the
formalism of Shklovskii \cite{shklpre}, in which the multivalent
counterions in the Stern layer are assumed to form a strongly
correlated liquid with short-range correlations resembling those
of a Wigner crystal. This theory provides a simple analytical
prediction for the charge-inversion concentration:
\begin{equation}
\label{eq:c0} c_0 = (\sigma_{\mathrm{bare}}/2erZ) \exp (-\mu_{c} /
kT) \exp(-\Delta\mu^0/kT).
\end{equation}
Here $\sigma_{\mathrm{bare}}$ is the bare surface charge density,
$\Delta\mu^0$ is the standard energy of adsorption of an ion and
$\mu_c$ is the chemical potential of the strongly correlated
liquid. The latter can be approximated by the value for a Wigner
crystal: $\mu_{c} \propto \sigma_{\mathrm{bare}}^{1/2}Z^{3/2}$. In
the calculations we use the full expression for $\mu_{c}$
\cite{shklpre}. In the absence of hydration effects and specific
chemical interactions, $\Delta\mu^0=0$ and $\mu_c$ is the sole
driving force behind charge inversion.

Qualitatively this theory with $\Delta\mu^0=0$ is in good
agreement with our observations. The predictions that charge
inversion is a general equilibrium effect and that $c_0$ depends
very sensitively on $Z$ but lacks dependence on the chemical
structure of the ions agree with our measurements.

A quantitative test of the theory is possible.
Equation~(\ref{eq:c0}) has two unknowns, $\sigma_{\mathrm{bare}}$
and $\Delta\mu^0$. From the consecutive measurements on
[Fe(CN)$_6$]$^{4-}$ and [Fe(CN)$_6$]$^{3-}$ in Fig~\ref{Figure3}
we extract values of $\sigma_{\mathrm{bare}}=+0.45$~e/nm$^2$ and
$\Delta\mu^0=1.4kT$, assuming that $\Delta\mu^0$ is the same for
both charge states of the ion. The corresponding values of
$\mu_{c}$ are 9.4 and 5.8$kT$ for $Z=4$ and 3, respectively. This
indicates that specific interactions are negligible and that ion
correlations are the dominant mechanism behind charge inversion in
this system.

The same calculation for the APTES measurements in
Table~\ref{Table1} yield values of
$\sigma_{\mathrm{bare}}=+0.2$~e/nm$^2$, $\Delta\mu^0=3.0kT$, and
$\mu_c=5.8$ and 3.5$kT$ for $Z=4$ and 3, respectively. This
suggests that in this case specific adsorption plays a larger
role. The difference between the two surfaces may occur because
the value of $\mu_c$ for APTES and $Z=3$ ions corresponds to the
lower end of the range of validity of Eq.~(\ref{eq:c0}). In
addition, the surface charge was modelled as being uniformly
distributed, whereas real surfaces consist of discrete chemical
groups; the relative importance of this disorder should be greater
for APTES with its smaller value of $\sigma_{\mathrm{bare}}$.

Taking $\Delta\mu^0=1.4kT$ and $c_0=200$~$\mu$M for
[CoC$_{12}$H$_{30}$N$_8$]$^{3+}$ screening silica gives
$\sigma_{\mathrm{bare}}=-0.4$~e/nm$^2$, in agreement with commonly
accepted values \cite{iler}.

These experiments are among the first systematic steps toward
understanding the fundamentals of screening of real surfaces by
multivalent ions. Specific binding does not provide an adequate
explanation for our observations. An alternative description based
on ion correlations provides qualitative and semi-quantitative
agreement with observations. In the future, measurements using
electrostatic gating will allow tuning the surface charge density,
permitting further quantitative tests of the theoretical
predictions.

\begin{acknowledgments}
We thank J. Lyklema for useful discussions and C. Dekker for
general support and useful discussions. This work was supported by
the 'Stichting voor Fundamenteel Onderzoek der Materie' (FOM) and
the 'Netherlands Organization for Scientific Research' (NWO).
\end{acknowledgments}



\begin{thebibliography}{21}
\bibitem[{not({\natexlab{a}})}]{note1}
\bibinfo{note}{Sum of metal ion radius and ligand (H$_2$O, NH$_3$, CN$^-$)
  diameter. The radii are comparable (within 4~\%) to crystallographic data}.
\bibitem[{sup({\natexlab{a}})}]{suppelec1a}
\bibinfo{note}{Supporting electrolyte was a 1~mM HEPES
  (4-(2-hydroxyethyl)piperazine-1-ethanesulfonic acid) buffer, pH~7.0$\pm$0.3
  set by adding KOH.}
\bibitem[{sup({\natexlab{b}})}]{suppelec1bc}
\bibinfo{note}{Supporting electrolyte was a mixture of 0.3~mM KOH and HCl with
  pH~6.5$\pm$0.5.}
\bibitem[{sup({\natexlab{c}})}]{suppelec3}
\bibinfo{note}{Supporting electrolyte was a mixture of 0.3~mM KOH and HCl with
  pH~5.8$\pm$0.3.}
\bibitem[{sup({\natexlab{d}})}]{suppelecmes}
\bibinfo{note}{Supporting electrolyte was a 0.1~mM MES
  (2-morpholinoethanesulfonic acid) buffer, pH~6.0$\pm$0.3 set by adding KOH.}
\end{thebibliography}
\end{document}